# Broadband Impedance Calculations and Single Bunch Instabilities Estimations of the HLS-II storage ring*

Zhang Qing-kun(张青鹍)[1)]　Wang Lin(王琳)[2)]　Li Weimin(李为民)[3)]　Gao Weiwei(高巍巍)[4)]

NSRL, School of Nuclear Science and Technology,
University of Science and Technology of China, Hefei 230029, P. R. China

**Abstract**: The upgrade project of Hefei Light Source storage ring is under way. In this paper, the broadband impedances of resistive wall and coated ceramic vacuum chamber were calculated using the analytic formula, the wake fields and impedances of other designed vacuum chambers have been simulated by the CST code, and then the broadband impedances model was obtained. Using the theoretical formula, longitudinal and transverse single bunch instabilities were discussed. With the carefully-designed vacuum chamber, we find that the thresholds of the beam instabilities are higher than the beam current goal.

## 1 Introduction

The Hefei Light Source (HLS) is a dedicated VUV and soft X-ray light source, where the beam energy is 800 MeV and the circumference of storage ring is 66.13m. To improve the quality of synchrotron radiation, an important Upgrade Project of HLS is undergoing, where the storage ring will be reconstructed. The main parameters of HLS-II storage ring are listed in table 1. Except for the lower beam emittance, the beam intensity will be higher than before, which is essential to the high synchrotron radiation flux and brightness[1].

Because geometric impedance-driven beam collective effects are of important considerations for reaching higher beam intensity, it is necessary to calculate the longitudinal impedances of components in the HLS-II storage ring, which may cause the longitudinal and transversal bunch instabilities and to estimate the acceptable beam current.

In this paper, short range wake fields of various vacuum chambers in the HLS-II storage ring are calculated by some analytic method and the CST code, and then the broadband impedance model is established. Based on the impedance model, single bunch instability thresholds, such as for the microwave instability, the transverse model coupling instability were estimated by the classical formulas [2~3].

## 2 Short review of Wakefield and impedance

When a charged particle is travelling through the storage ring, if the wall of the beam pipe is not perfectly conducting and having some discontinuities, the movement of the image charges will be slowed down, thus leaving electromagnetic fields behind, which will influence the motion of the later particles. So we calculate the longitudinal and transverse wake functions by integrating over the EM force normalized by the charge $q$ and a horizontal offset $\xi$ [3~5]:

$$\begin{cases} W_{\parallel}^{'} = -\frac{1}{q}\int_{-\infty}^{\infty} dz E_z. \\ W_{\perp} = \frac{1}{q\xi}\int_{-\infty}^{\infty} dz \left(\vec{E}+\vec{v}\times\vec{B}\right)_{\perp}. \end{cases} \quad (1)$$

Table.1 Parameters of HLS-II storage ring

| | |
|---|---|
| Beam Energy [MeV] | 800 |
| Circumference [m] | 66.13 |
| Momentum Compaction Factor | 0.02 |
| Transverse Beam Emittance [nm-rad] | 37 |
| Natural Energy Spread | 0.000472 |
| Nominal RMS Bunch Length [mm] | 14.8 |
| Vacuum Chamber Height Size [m] | 0.04 |
| Synchrotron Frequency [kHz] | 28.0 |
| Betatron Tunes | (4.4141, 3.2235) |
| Average Transverse Beta Function [m] | 8.50/5.25 |
| Energy Loss per turn [keV] | 16.73 |
| Main RF Frequency [MHz] | 204 |
| Single Beam Current $I_{dc}$ [mA] | 300/45 |
| Damping Time $\tau_x / \tau_y / \tau_z$ [ms] | 23/22/10 |

---
\* Supported by the Natural Science Foundation of China (11175182 and 11175180)
1) E-mail: zhangqk@mail.ustc.edu.cn
2) E-mail: wanglin@ustc.edu.cn
3) E-mail: lwm@ustc.edu.cn
4) E-mail: gaoww@ustc.edu.cn
Submitted to 'Chinese Physics C'

For a Gaussian bunch, the wake potential can in principle be found from the convolution of the wake function with the normalized line density $\lambda(\tau - t)$.

Actually, we often use the coupling impedance for analytical study of the beam instabilities, which is the Fourier transformation of the wake function. We also use the reduced impedance for the longitudinal beam instabilities, which is a comparable quality for storage rings with different circumference. The reduced impedance is defined as the impedance divided by the harmonic number $n = \omega/\omega_0$.

## 3 Wake fields and impedances for HLS-II storage ring

In the HLS-II storage ring, we can ignore the influence of space charge effect due to the highly relativistic beam. In this section, we considered the resistive wall and the discontinuities of vacuum chambers by analytic formula and the CST code[6]. The wake potentials were calculated up to s=100mm, where s is the bunch coordinate. We used a Gaussian bunch with the rms bunch length $\sigma_z$ of 3mm and the charge $Q$ of 0.0000001C. In the wakefield solver, all metallic parts were considered to be perfect electric conductors, and open boundary conditions are assigned at the entrance and exit of the structure.

### 3.1 Resistive wall

The vacuum chamber's octagonal cross-section of the HLS-II storage ring was shown in Fig 1, including its inner dimensions. Almost all of the chambers are made of Stainless Steel (SS).

The impedance of the resistive wall is calculated analytically. In order to estimate the resistive wall impedance, a simplified modal was introduced. Here all the cross section of the vacuum chamber is considered to be elliptic, which has a length of major axis $2a$ and minor axis $2b$ shown in Fig 1.

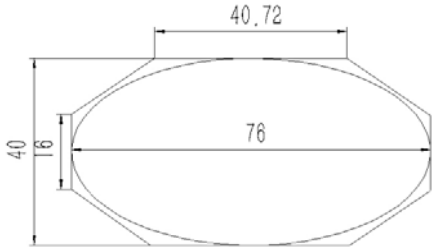

Fig.1. The cross-section of the SS vacuum chamber. All the dimensions are in unit of mm. The pipe shape indicated by the ellipse was used in the impedance estimation.

$$a = d \cosh u_0, \quad b = d \sinh u_0, \quad d^2 = a^2 + b^2 \quad (3)$$

The longitudinal resistive wall impedance of this elliptic pipe is given by: [7~8]

$$\frac{Z_{//}}{n} = Z_0 \frac{(1-i)}{2} \frac{\delta}{b} \frac{L}{2\pi R} G_0(u_0). \quad (4)$$

Where

$$G_0(u_0) = \frac{\sinh u_0}{2\pi} \int_0^{2\pi} \frac{Q_0^2(\upsilon) d\upsilon}{[\sinh^2 u_0 + \sin^2 \upsilon]^{1/2}}.$$

$$Q_0(\upsilon) = 1 + 2\sum_{m=1}^{\infty} (-1)^m \frac{\cos 2m\upsilon}{\cosh 2m u_0}.$$

Based on the Panofsky-Wenzel theorem, the transverse resistive wall impedance of the elliptic pipe can be estimated with the effective radius $c$:

$$Z_\perp = \frac{2R}{c^2} \frac{Z_{//}}{n}. \quad (5)$$

In this case, the conductivity of SS beam pipe is $1.5 \times 10^6 (\Omega m)^{-1}$. Then we can obtain the impedances of the resistive wall:

$$\begin{cases} \dfrac{Z_{//}}{n} = 4.5212(1+j)/\sqrt{n} \; \Omega \\ Z_x = 33.865(1+j)/\sqrt{n} \; K\Omega \\ Z_y = 36.8767(1+j)/\sqrt{n} \; K\Omega \end{cases} \quad (6)$$

### 3.2 Coated ceramic vacuum chamber

In HLS-II storage ring, there are four coated ceramic vacuum chambers (fig2) inside the kicker magnets. A metal coasting is required on the inner surface of the ceramic vacuum chamber, the thickness of which will affect the pulsed magnetic field of the kicker magnet and the coupled impedance of the ceramic vacuum chamber. With carefully design, the thickness of the metal coasting is $2\mu m$ in the ring. For this kind of structure, it is very difficult to calculate the coupling impedance by the CST code, because of the small thickness of the metal coating. Hence a simple model of an infinitely long pipe was used to estimate the coating impedance.[8]

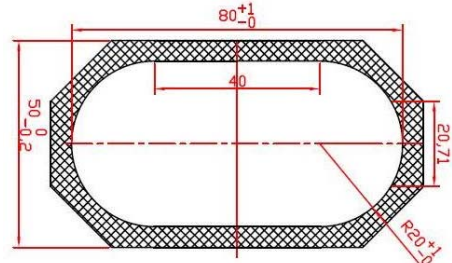

Fig.2. The cross section of ceramic vacuum chamber of the injection system at HLS-II

The longitudinal impedance formula is presented below:

$$\frac{Z_{//}}{L} = Z_{met} \frac{A + \tanh(\kappa d_m)}{1 + \tanh(\kappa d_m)}. \quad (7)$$

Where

$$Z_{met} = \frac{1 - i\,\text{sgn}(\omega)}{bc}\sqrt{\frac{|\omega|}{2\pi\sigma}}\frac{1}{\sqrt{4\pi\varepsilon_0}}.$$

$$A = \frac{(1 - i\,\text{sgn}(\omega))d_c}{d_s}\left(1 - \frac{1}{\varepsilon}\right).$$

$$\kappa = \frac{1 - i\,\text{sgn}(\omega)}{\delta}.$$

$$d_s = \frac{c}{\sqrt{2\pi\sigma|\omega|}}\sqrt{4\pi\varepsilon_0}.$$

In HLS-II, the length of this piece $L$ is 0.28m, the thickness of the metal layer $d_m$ is $10^{-6}$ m, the thickness of the ceramic chamber $d_c$ is 0.005m, and the equivalent radius of the vacuum chamber is 0.2m, then we obtain that: $Z_{//} = 0.002497n + 0.014774j$.

### 3.3 BPMs

There are 36 BPMs in the HLS-II storage ring. Each BPM consists of four button-type electrodes, which are located on the upper and lower planes of the vacuum chamber without electrical connection. The radius of the electrode is optimised to avoid the HOM modes being trapped in the cavity like.

To estimate the transverse and longitudinal impedances, the designed BPMs were modelled on the octagonal cross-section for HLS-II. A high resolution hexagonal mesh is required for accurate discretization of many complex details with sizes < 1 mm. Fig. 3 shows the longitudinal wake potentials of the BPM computed over a distance of 0.1m at 10 lines per wavelength (in terms of CST parameters), but with two different constraints for the maximum longitudinal mesh step: the red and blue wake potentials correspond to a maximum longitudinal mesh step of 0.1mm (250 million mesh cells per quarter of the 3D structure) and 0.01mm (2500 million mesh cells per quarter of the 3D structure) respectively. As expected, a mesh with a smaller longitudinal step results in a better accuracy for the wake field solution: in particular, there is a significant reduction in the unphysical wake occurring ahead of the bunch.

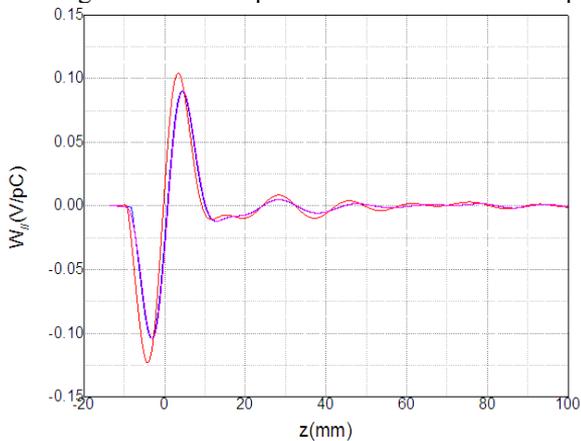

Fig.3. Longitudinal wake potential of BPM computed by CST(red and blue lines); wake potential convoluted from the wake function(magenta line)

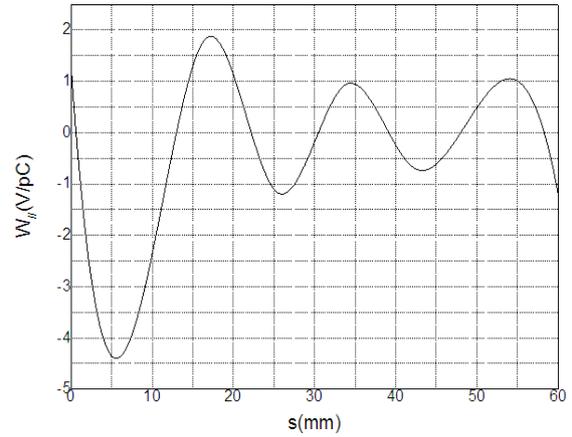

Fig. 4: Longitudinal wake function of the BPM

We can find that the wake potentials computed directly by CST are compared with the wake potentials evaluated as a convolution and the wake function (Fig. 4): the results are in good agreement, which indicated that the function has been computed with reasonable accuracy.

### 3.4 Pump ports

In order to reduce the coupling impedance of the pump port, we made the pump port with a mesh aperture design to meet the high pumping speed and the continuity of the vacuum chamber in the HLS-II storage ring. There are four kinds of pump ports, their longitudinal impedance, the loss factor and the kick factor were calculated by the CST code.

### 3.5 Longitudinal feedback kicker

In order to damp the longitudinal coupled-bunch instabilities, we designed and installed a broadband longitudinal feedback kicker, which is a waveguide overloaded pillbox cavity with two input and two output ports. Meanwhile, there are two nose cones introduced to improve the shunt impedance of the cavity. This method leads to a strong coupling between the pillbox and the waveguides, which can help damp the harmful HOMs. In the calculation, we treated the interface of $x_{min}$ $x_{max}$ $y_{min}$ and $y_{max}$ as an outgoing wave guide port. Fig. 5 shows the longitudinal wake potential computed for the longitudinal feedback kicker, compared the wake potentials in the BPM and the longitudinal feedback kicker, we can find that the later particle will suffer larger effects in the later component.

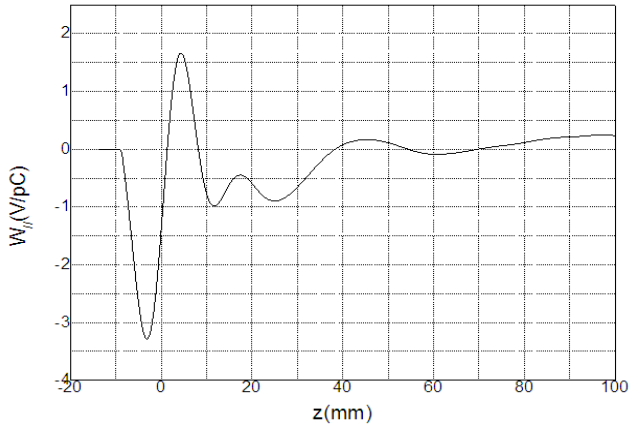

Fig.5. Longitudinal wake potential of longitudinal feedback kicker

### 3.6 Clearing electrodes

In the HLS-II storage ring, we designed the DC cleaning electrode to overcome the ion trapping, which is one of serious collective effects in the low-energy storage ring. The clearing electrodes have a length of 525 or 615mm, which have been built in the vacuum chamber with four ceramic material supports and a high voltage electrode. Thus, the full-height of the vacuum chamber is 43.5mm, which is 3.5mm higher than the common vacuum chamber.

### 3.7 RF cavity

At present, there is an old RF cavity installed in the HLS-II storage ring, which is an important source of impedance. The 3D model of the RF cavity is built by the CST code. Its length is 0.472m. Only the fundamental mode of the cavity has a high quality factor and impedance, while all the other modes have low quality factors due to energy loss, which was discussed in the other paper. Fig. 6 shows the longitudinal wake potential computed for the RF cavity, we can find that, the wake potential also have a large value after a long distance.

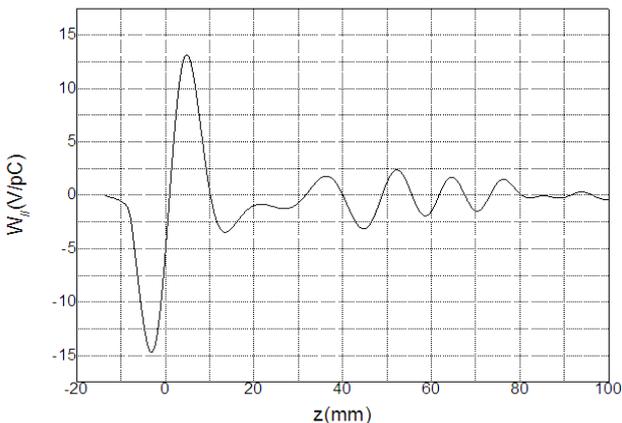

Fig.6. Longitudinal wake potential of RF cavity

### 3.8 Curved section

In a bending chamber, an electron beam moves along the curvilinear trajectory. Simulation of a curvilinear structure is impossible in CST. For this reason, a simplified rectilinear model of the HLS-II bending chamber including the octagonal beam channel, the pumping ports, and the BPM has been created (fig.7). Analysis of eigenwaves performed with CST shows that there is no longitudinal resonant mode excited in the structure.

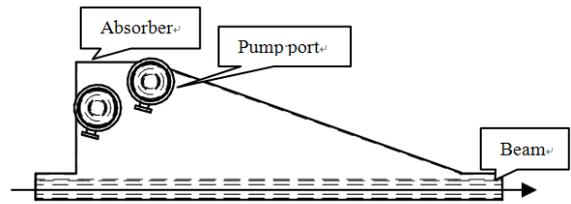

Fig.7 the curved section

### 3.9 Others

In addition to these main components, there are a large number of small discontinuities in the HLS-II storage ring, such as the connection between the flange and the vacuum chamber, shielded bellows, slots for synchrotron radiation monitor. In this paper, we used the CST code to set models and calculate the detailed wake fields and impedances. The result of which is shown in Fig8.

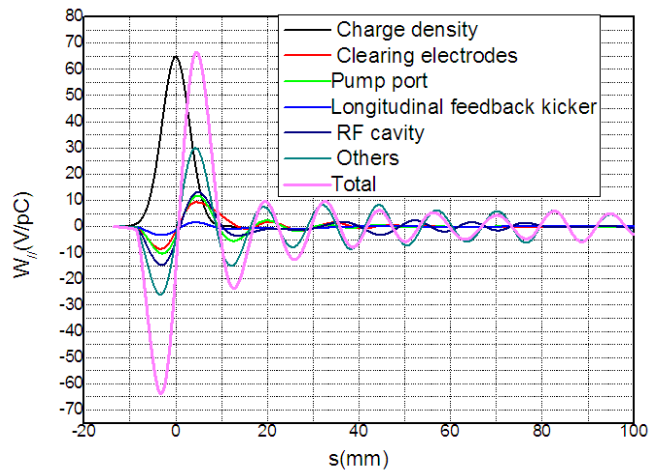

Fig8. Longitudinal wake potentials as a function of the distance from the bunch head for various groups of elements.

From the Fig 8, we can find that the most significant wake potentials are produced by the pumping ports, the clearing electrodes, the longitudinal feedback kicker, the RF cavity and the curved section.

Table 2: calculated impedance for components of the HLS-II storage ring

| Component | N | $|Z_{//}/n|$ Ω | k V/pC | $k_x$ V/pC/mm | $k_y$ V/pC/mm |
|---|---|---|---|---|---|
| RF cavity | 1 | 0.018408 | 1.620064 | 2.361957 | 3.755356 |
| Pumping port | 25 | 0.0046 | 1.3888 | 25.5577 | 30.9911 |
| Flange | 59 | 0.2548 | 3.4787 | 1.8788 | 1.5853 |
| BPM | 36 | 0.0036 | 0.4301 | 1.6607 | 1.7164 |
| Clearing electrodes | 32 | 0.0849 | 3.4887 | 72.4219 | 77.0221 |
| bellows | 16 | 4.6240e-05 | 0.1313 | 3.4647 | 3.4486 |
| Longitudinal feedback kicker | 1 | 0.0095594 | 8.347e-001 | 8.221e-001 | 8.434e-001 |
| Curved section | 8 | 0.4435 | 27.6218 | 18.6493 | 45.3778 |
| Others | - | 1.5517e-04 | 0.6 | 2.8964 | 2.7163 |
| Total | 178 | 0.819569 | 39.59416 | 129.7136 | 167.4564 |

## 4. Impedance budget

The impedance budget[9~11] of the HLS-II storage ring is shown in Table 2, where the longitudinal broadband impedance, the longitudinal loss factors, and the transverse kick factors are presented for each element. The major contributors to the impedance of the storage ring are the RF cavity, the pumping port chambers, the longitudinal feedback kicker and the flanges. The results indicated that the total longitudinal broadband impedance is about 0.82 Ω. Then we can obtain the broadband impedance model of the HLS-II storage ring as:

$$Z_{//}/n = 0.819569j + \frac{4.5212(1+j)}{\sqrt{n}} + \frac{0.014774j}{n} + 0.002497. \quad (8)$$

Where the first term of the formula is the broad-band impedance at low frequency, the second term is the resistive wall impedance, and the last two terms are the impedance of the coated ceramic vacuum chamber.

## 5. Instability threshold estimation

In this section, the total value of the longitudinal broad-band impedance was used to study the single bunch instabilities in the HLS-II storage ring.

### 5.1 Microwave Instability

When the peak current of a single bunch is higher than a threshold current, the bunch lengthening and the energy spread increasing are generated until the peak current is reduced to the current threshold. Meanwhile, the microwave instabilities [12] also affect the brightness of the insertion device:

$$N_{th} = \frac{(2\pi)^{3/2} \alpha_c \sigma_\varepsilon^2 (E/e)}{ce} \frac{\sigma_z}{|Z_{//}/n|}. \quad (9)$$

Where the impedance $|Z_{//}/n|$ is calculated from the wake field using the CST code, $\sigma_z$ is the rms bunch length, $\sigma_\varepsilon$ is the natural energy spread, and then the value $N_{th}$ given by the equation (9) has been found to be $2.1637e+10$. Then we can find that in the HLS-II storage ring, the threshold of single bunch current for the microwave instability is 15.7mA. Even considering the setting up error and some components missing in the estimation, we can find that the threshold of single bunch current for the microwave instability is larger than the designed value.

### 5.2 Transverse mode coupling instability

When the beam current is increased, the coherent frequencies of head-tail modes are shifted and some of them may be coupled, which results in a fast beam instability. This is called transverse mode coupling instability. For Gaussian bunches and the broadband resonator impedance, $I_{thresh}$ can be expressed with kick factor $\kappa_\perp(\sigma_z)$ which eliminates the need for a bunch length correction factor: [12]

$$I_{thresh} = \frac{C_1 f_s E/e}{\sum_i \beta_i \kappa_{\perp i}(\sigma_z)}. \quad (10)$$

The constant $C_1$ is about 8. Using the numerical value in Table 1, we can obtain the single bunch current threshold is 12.589A, which is much larger than the one for the longitudinal microwave instability. For a more exact calculation of the threshold we should use computer programs solving for the coherent modes, including the bunch lengthening and the potential-well deformation with the beam current. Nevertheless, we can conclude that the longitudinal microwave instability is much more significant in the storage ring.

## 6. Conclutions

The HLS-Ⅱ storage ring broad-band impedance and loss factor have been calculated. Contributions to the total broad-band impedance from different groups of elements have been found. Meanwhile we estimate the instabilities based on the broadband impedance model, which did not affect the beam quality seriously. In future, we will use the broadband impedance to study the beam instabilities by tracking, and then we will calculate the narrowband impedance and study the coupled bunch instabilities for the HLS-Ⅱ storage ring.